\documentclass[a4paper,twoside]{article}
\newcommand{\affil}[1]{$^{\rm #1}$}
\usepackage[authoryear]{natbib}
\bibpunct{(}{)}{;}{a}{}{,}
\usepackage{graphicx}
\date{} 
\newcommand{\kms}{\mbox{km\,s$^{-1}$}}

\title{\large\bf\flushleft 
The rebrightening of planetary nebulae through ISM interaction}
\author{\parbox{\textwidth}{\flushleft
\vspace{-0.5cm}
{\it C.~J. Wareing\affil{A,B}}\\
\vspace{0.4cm}
{\small \affil{A}\,Department of Applied Mathematics, University of Leeds, Woodhouse Lane, Leeds, LS2 9JT, UK.}\\
{\small \affil{B}\,E-mail: cjw@maths.leeds.ac.uk}}}
\begin{document}
\twocolumn[
\begin{changemargin}{.8cm}{.5cm}
\begin{minipage}{.9\textwidth}
\vspace{-1cm}
\maketitle
%
%
\small{\bf Abstract:}

The interaction of planetary nebulae (PNe) with the interstellar medium
as they move through it is now acknowledged to be a major shaping
effect not just for ancient and large PNe, but also for relatively young 
PNe with high speed central stars. The most common effect is a rebrightening 
as the PN shell interacts with a pre-existing bow shock structure formed
during the previous evolutionary phase of the central star. In this review,
we consider this rebrightening in detail for the first time and discuss 
its origins, highlighting some observed
examples. We go on to discuss the AGB star progenitors, reviewing the 
evidence for bow shock structures, and consider the progeny of rebrightened 
PNe - strongly disrupted objects which bear very little resemblance to
typical PNe. Sh 2-68 is inferred to be perhaps the only documented case
so far of such a PN.

\medskip
\medskip
\end{minipage}
\end{changemargin}
]
\small

\section{Introduction}

Using simple hydrogen atoms, stars forge the entire range of atomic 
species we observe in Nature. Towards the end of their lives, stars 
eject these newly forged atoms into interstellar space. Out of this 
enriched interstellar gas, new stars, their planets and even life 
form. Stellar mass loss is the motor that makes stars and galaxies 
change over cosmic time. Yet, when and how mass is lost from stars 
is still an open debate. Stars more than eight times more massive
than the Sun end their lives in spectacular explosions known as
supernovae. Stars less than eight times more massive than the Sun 
lose up to 90\% of their mass when they become giants, shortly after 
running of out hydrogen in their cores. The mass that is lost 
eventually forms beautiful nebulae around the mother star, objects 
that we call planetary nebulae; a misnomer acquired 
because they looked like the small circles of planets to 17th century 
astronomers. Modern telescopes have revealed they exhibit 
complex shapes (butterfly, multiple lobes emerging from disks, jets 
and bullets, objects pretty enough to fill coffee table books!). These 
shapes are in need of an explanation: how do stars lose 
quite so much mass and why is this mass not residing in a more or less 
spherical distribution around the star?

Simple round shapes have been understood in terms of the Interacting 
Stellar Winds (ISW) model \citep{kwok82}. In this model, planetary 
nebulae (PNe) are formed when stellar material surrounding the central star (CSPN),
originating in a slow wind from the asymptotic giant branch (AGB) 
progenitor, is swept up into a nebular shell by a hot, fast, ionizing
post-AGB wind from the white dwarf CSPN. The great majority of PNe are
not spherical and many theories have been introduced to explain the wide 
variety of observed shapes. Many shapes can be reproduced in models by
introducing an asymmetry into the slow wind \citep{kahn85,balick87}.
Later theories have concentrated on the origins of the asymmetry, 
considering stellar rotation and/or magnetic fields 
\citep{garcia-segura99,frank04,garcia-segura05}. Most recently though,
such effects have been shown to be difficult to maintain in single
AGB stars, leaving binary central star systems as the primary progenitors
of PNe (see, for example, \cite{demarco09} for a full review and 
discussion of the binary progenitor hypothesis).

There are several cases though where only the outer shells show a
departure from symmetry. This has been postulated to arise from an 
interaction with the interstellar medium (ISM). In this review, we will 
consider the current knowledge in this area and the previous work which 
has led up to the interpretation of the PN--ISM interaction as a four 
stage evolutionary process, including a period of 'rebrightening', 
typically occurring late in the lifetime of a PN. We will review this 
particular part of the process, including the dominating effect of the earlier AGB phase of
evolution, and consider the immediate progeny of rebrightened PNe, 
objects which can bear little resemblance to their origins.
The observational support for each stage will also be summarized here
for the first time.
Finally, we will discuss how rebrightening changes the lifetime
estimates of typical PNe and how a sizable sample of interacting
PNe may change estimates of the PN population and the understanding
of ISM structure on a galactic scale.

\section{Understanding the interaction of PNe with the ISM}

The idea that a PN would interact with its surroundings as it moved 
through the ISM was first extolled by \cite{gurzadyan69}. \cite{smith76}
performed the first theoretical study in which a thin shell approximation
was assumed; the 'snow-plough' momentum-conserving model of \cite{oort51}
was applied to shape the external shell of the PN. \cite{isaacmann79} 
used the same approximation with higher relative velocities and ISM 
densities. Both concluded similarly: that a PN would fade away before
any disruption of the shell by interaction with the ISM became
noticeable.

The work of \cite{borkowski90} challenged this standpoint. They
investigated CSPN with proper motions greater than 0.015 arcsec\,yr$^{-1}$
and found that all of the PNe showed signs of interaction with the ISM,
including dipole asymmetries, distortions in the direction of the
CSPN motion and displacement of the CSPN from the geometric centre of
the nebula. At the same time investigating PNe with large angular extent, they 
found many of these also showed signs of interaction. Concluding that
PN--ISM interaction must be commonplace, \cite{borkowski90} suggested
a simple evolution whereby the PN shell is first compressed in the
direction of motion and then in later stages this part of the shell
is significantly decelerated with respect to the central star. They
calculated that the
interaction would become apparent when the densities in the expanding
shell dropped below $n_{\rm H} = 40\ {\rm cm}^{-3}$ for a PN in the Galactic Plane.
As the PN fades, only the brightest regions where the shell is 
compressed remain detectable.

\cite{soker91} performed the first hydrodynamic simulations modelling 
the interaction, starting with the nebula shell already formed but
above the upper density limit for ISM interaction to become apparent
derived by \cite{borkowski90}. They validated the thin shell approximation for
calculating the displacement of CSPNs from the geometric centres of
their nebulae and concluded similarly that the interaction with the
ISM becomes noticeable only once the density of the shell drops below the
same critical limit.

The conclusions of \cite{borkowski90} and \cite{soker91} reconciled
the theory and observations, allowing for ISM interaction to become
apparent once a PN had become large enough for the densities in the
shell to have dropped below the critical limit, but {\it before} it faded
away. At this stage though, ISM interaction was not thought to cause
any increase of emission at late stages (i.e. a rebrightening),
but instead to cause some areas - those interacting with the ISM -
to fade more slowly than the rest of the PNe.

\cite{soker91} also noted that this simple picture breaks down for high 
velocity PNe in low density environments. Here, a Rayleigh--Taylor (RT) 
instability develops, leading to shell fragmentation. This fragmentation 
was considered in detail by \cite{dgani94} and then \cite{dgani98}. In 
these papers, the authors performed hydrodynamic simulations of the
RT instability and found that it plays an important role in the 
shaping of the outer regions of PN surrounding high-speed central stars. 
They suggested the RT instability can cause fragmentation of the 
nebular shell in the direction of motion. Any fragmentation caused by 
this instability would only be present if the CSPN relative velocity 
was greater than 100 \kms. They also noted the situation could also be 
further complicated by the Kelvin-Helmholtz (KH) instability and
magnetic fields, if inclined to the direction of motion, could break
the cylindrical symmetry of the process and elongate the fragmentary
structures \citep{dgani97}.

Consequently, the field reached a consensus that interaction of PNe 
with the ISM becomes observable only during the late stages of PN 
evolution and the search for
evidence of the interaction became restricted to PNe with 
large angular extent \citep{tweedy96,xilouris96,kerber00,rauch00}
where it was commonly found and such PNe were classed as
ancient. The definitive Atlas of Ancient PNe 
\citep{tweedy96} found observational evidence for ISM interaction in 
21 of their 27 targets.

In the first work heralding the next phase of PN--ISM modelling,
\cite{villaver03} pointed out that the interaction had previously been 
studied by considering the interaction only after the nebular shell had 
formed. Villaver et al. performed 2D hydrodynamic simulations following the 
preceding AGB phase of evolution, including the latest theoretical 
model of mass-loss variation, and found that crucially the 
interaction begins during that phase when the slow AGB wind is shaped 
by the ISM. The PN forms much later, initially cocooned within a bubble of
undisturbed-by-ISM AGB wind material which, when the CSPN is moving 
through the ISM, is bounded by a bow shock formed by the AGB wind-ISM 
interaction and located at the point of ram pressure balance between 
the slow wind and the oncoming ISM. Choosing a conservative relative 
velocity for the CSPN, $v_{\rm ISM} = 20$ \kms\ and a low density of the 
surrounding ISM of $n_{\rm H}=0.1$ cm$^{-3}$, they showed that PN--ISM
interaction can become apparent at a relatively middle age with low velocities and 
low ISM densities. They also found the first evidence for a 
rebrightening, occurring when the PN shell has expanded far enough to
interact with the AGB wind-ISM bow shock, compressing the shell in the 
direction of motion and increasing the density and temperature and thus the brightness. In their
low-speed, low-density simulation though, this still occurred at a middle to late
stage in the PN lifetime.

Showing that neither high velocities, high ISM densities, nor the presence of a 
magnetic field were necessary to explain the observed asymmetries in 
the external shells of PNe, \cite{villaver03} allowed the PN--ISM
interaction to be applied to the majority of PN, not just an exceptional
few. This process became important for all PNe.
Their simulations also hinted that ram-pressure-stripping of mass
downstream during the AGB and postAGB/PN phases may provide a possible 
solution to the problem of missing mass in PN whereby only a small 
fraction of the mass ejected during the AGB phase is inferred to be 
present during the post--AGB phase.

At the time, observational evidence for the effect of the ISM on AGB 
wind structures was limited. \cite{zijlstra02} presented evidence for
a detached dust ring surrounding an AGB star, which at 4pc in diameter
was unrivaled in size. They noted that such detached shells are 
normally interpreted in terms of late-AGB stage thermal pulses, however
in this particular case they concluded a significant proportion of the
shell may consist of swept-up ISM, with the detached appearance 
explained by the wind-ISM interaction.

Motivated by recent observations of the PN Sh\,2-188 which revealed the
true nature of the extended faint emission behind the well-known bright arc of the
nebula, of which we will say more later as it is a useful example
of rebrightening in PNe, \cite{pn-ism} developed a triple-wind model
in the frame of reference of the CSPN, 
including an initial slow AGB wind, a subsequent fast post--AGB wind, 
and a third continuous wind reflecting the movement through the ISM.
Employing a 3D hydrodynamic scheme, they performed a range of 
simulations investigating the effect on PN formation of relative 
velocity, ISM density and CSPN mass-loss rate during the AGB and pAGB. 
With this extensive investigation of parameter space, they were able
to generalise a four stage evolutionary classification of PN--ISM 
interaction, driven by the process of ram-pressure-stripping. In the next section
we review the support for the idea of PN rebrightening in the context of the first three
stages of this classification.

\section{Rebrightening as part of the four stages of PN--ISM interaction}

\begin{figure}[ht]
\begin{center}
\includegraphics[angle=0,width=7cm]{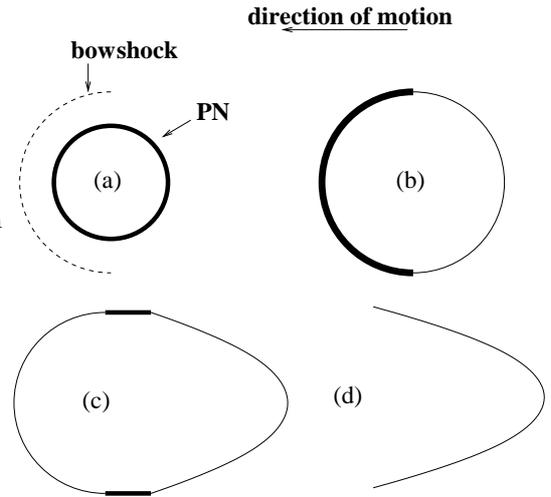}
\caption{A reproduction of figure 7 from \cite{pn-ism} showing a simple 
illustration of the appearance of a PN during the four stages of PN--ISM 
interaction. The direction of motion is to the left and thicker lines 
indicate the brightest regions. Panel (a) illustrates stage WZO 1, (b) 
stage WZO 2), (c) stage WZO 3 and (d) stage WZO 4. The position of 
letters (a), (b), (c) and (d) indicate the position of the central star 
at each stage.}
\label{pnstages}
\end{center}
\end{figure}

In order to demonstrate how a PN can be rebrightened, we will review
the first three stages of PN--ISM interaction, according to the 
classification of \cite{pn-ism}. In the first stage,
the PN expands inside the cocoon formed by the AGB 
wind-ISM bow shock formed during the preceding AGB phase, unhindered by 
any ISM interaction. The AGB wind-ISM interaction is radially further 
from the central star and it is possible to observe a faint arc of the 
bow shock around the PN. Such an arc is seen in several cases including
the famous Dumbbell nebula, where clumps of material surrounding the 
bright nebula are inside what appears to be AGB wind-ISM bow shock 
\citep{meaburn05}. Early models of the interaction which
did not take account of the AGB-ISM interaction could not explain the
presence of this structure; an explanation required the insight of 
\cite{villaver03}. In the case of a slow-moving star with a radially
distant bow shock, Wareing et al. noted this stage can last for 
the entire PN lifetime and hence a PN--ISM interaction would never be 
observed.  However, if the central star is moving even at average speed 
through the ISM (i.e. $\sim 50$ \kms \citep{burton88}), then in support
of \cite{villaver03}, they noted PN--ISM interaction can rapidly
occur. In some extreme cases with low CSPN mass-loss rates in high density 
regions, Wareing et al. noted that interaction can appear at a very
young age of only 1000 years. We reproduce figure 7 from 
\cite{pn-ism} here in Figure \ref{pnstages} to illustrate the primary 
characteristic of this stage, a swept-up ISM shell up to a few pc away.

The PN enters the second stage when it has expanded far enough to 
interact with the AGB wind-ISM bow shock. This is the point at which 
rebrightening occurs. As the expanding PN shell encounters the AGB wind 
bow shock in the direction of motion, it merges and drives another shock 
through the bow shock, increasing the density and temperature of the 
mixed material in that region and strengthening the emission, a 
rebrightening, the primary characteristic of this stage. This is 
illustrated in Figure \ref{pnstages}(b). Wareing et al. note that if 
the motion is predominantly in the plane of the sky, this part of the 
nebular shell may be brighter than the rest, although the PN will 
continue to appear circular on the sky. As the appearance deviates 
from circularity/symmetry and the shell is decelerated in the direction
of motion by the AGB wind-ISM bow shock, rather than just interaction
with the ISM as originally thought by \cite{borkowski90} and 
\cite{soker91}, Wareing et al. define the PN to be entering the third 
stage of interaction.

This stage is characterized by the shift of geometric centre of the 
nebula downstream away from the CSPN as illustrated in Figure 
\ref{pnstages}(c). The shift, due to the deceleration of the PN 
shell in the direction of motion, is guaranteed to occur and provides a 
measurable effect of the interaction. Again, this has been noted by several 
authors \citep{borkowski90,soker91,villaver03,pn-ism}. 

The first three stages build the model which accounts for many of
the characteristics of PN--ISM interaction noted by previous authors,
e.g. \cite{borkowski90,tweedy96}. In contrast to previous models though,
the one-sided brightness bias observed appears to be a consequence of
rebrightening, and is followed by deceleration of the shell in the direction
of motion and shift of the CSPN away from the geometric centre of the
nebula. We will now go on to discuss several cases of this situation,
beginning with Sh\,2-188 which could be considered as a template
for future studies of rebrightened PNe.

\section{Observed cases of rebrightened PN}

\begin{figure}[ht]
\begin{center}
\includegraphics[width=7cm]{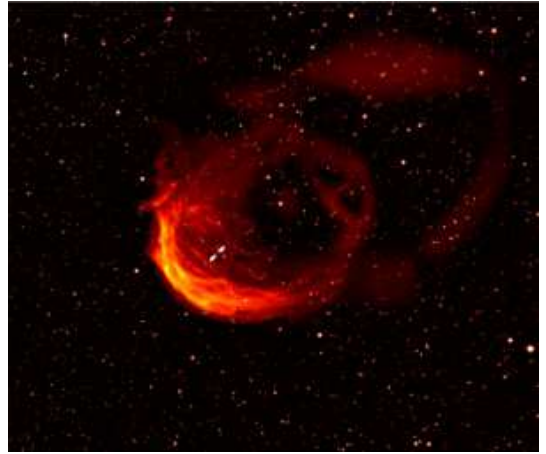}
\caption{A reproduction of figure 3 from \cite{sh2-188} showing a
combination of observations of the PN Sh\,2-188 taken as part of IPHAS. 
The faint central star is indicated on the image between the markers.
In the image, North is up and East to the left.}
\label{pn-sh2-188}
\end{center}
\end{figure}

In the Atlas of Ancient PNe \citep{tweedy96}, Sh 2-188 was remarked
upon as a bright, one-sided filamentary arc-like PN with 
faint, rather homogeneous material opposite the bright arc revealed by
deeper H$\alpha$ images. Whilst
commenting on the possibility of this material being a comet-like tail,
the authors opted for the simpler explanation that Sh 2-188 resides in
a highly inhomogeneous ISM. 2003 observations, taken as part of the 
Isaac Newton Group Photometric H$\alpha$ Survey of the Northern 
Galactic Plane (IPHAS) \citep{drew05} revealed the faint material to be
a ring-like completion of the arc and a tail stretching away in 
opposition to the bright arc. \cite{sh2-188} modelled the nebula
as a strong PN--ISM interaction where the CSPN is moving at 125 \kms\ 
in the direction of the bright arc relative to the ISM. In Figure 
\ref{pn-sh2-188}, we reproduce their figure 3 combining the IPHAS 
observations of the nebula. The nebular 
shell is interacting with the AGB wind-ISM bow shock and undergoing
the exact rebrightening discussed in the previous section. Sh\,2-188
is a prime example then of the effect under discussion in this review. The faint completion
of the arc is the original nebular shell continuing to expand into
AGB wind material cocooned by the bow shock and tail. The tail consists
of far-older ram-pressure-stripped material from the head of the bow
shock stretching away to the North-West.

Wareing et al. note that if the central star had not been traveling through the local ISM, 
Sh\,2-188 would most likely have a ring-like appearance on the sky with 
a radius similar to the distance between the faint completion of the
arc and the CSPN - approaching twice the actual radius.
The surface brightness of the whole object would
be comparable to the faint completion of the ring. 
Clearly, ISM interaction has strongly affected the late evolution of
this particular nebula, limiting its expansion and rebrightening the
nebula on the upstream side of the shell. Finally, \cite{sh2-188}
performed a proper motion analysis and found the CSPN to have a 
considerable motion of 0.03 +- 0.01 arcsec\,yr$^{-1}$ in the direction
of the bright arc, confirming the PN--ISM interaction hypothesis. The
combination of the proper motion and the high velocity required to
model the appearance of the nebula, placed it at distance of 850 pc,
consistent with other estimates of the distance to the star.

The 21 candidate PN--ISM interactions in the Atlas of Interacting PNe appear to show
similar rebrightening characteristics. Of those, Sh 2-216 shows
filamentary evidence of a particularly complex magnetized ISM in the vicinity of the
PN \citep{ransom08}. Several objects in the search for extensive halos
around PNe \citep{haohsia} also show evidence for one-sided rebrightening,
suggesting the interaction is commonplace. There is also the complex case
of the bipolar nebulosity surrounding the nova GK Per: 
thought to be an ancient PN ejected by a member of the binary system,
the nebula has a swept back appearance, brightened on the upstream
side, in alignment with the proper motion of the system \citep{bode04}.

Evidence for other PNe with tails is considerably rarer. Deep H$\alpha$ 
+ N{\sc II} observations of HFG1 \citep{xilouris96,boumis09} reveal a strong
candidate tail. \cite{szent} find the emission of [S {\sc II}] in the PN 
NGC 246 probably traces the interaction of the PN with the ISM. \cite{meaburn00}
find mid-IR emission apparently tracing a tail behind NGC 3242, supported
by \cite{ramos09} who find evidence for ram-pressure stripped material
in two other PNe NGC 2440 and NGC 6629. Finally, 
the complex and asymmetric ancient PN Sh 2-68 appears to have evidence
for a long tail, stretching away from the location of the CSPN in 
opposition to the direction of proper motion \citep{xilouris96}. This object
may be the first documented case of the progeny of a rebrightened PN, 
but first we will discuss the characteristics of the AGB progenitors, and
the evidence in the literature for such objects.

\section{The progenitors of rebrightened PNe}

The AGB progenitors of ISM-interaction-rebrightened PNe must be moving 
through the ISM at considerable velocities. \cite{sh2-188} hypothesized
that bow shock-like structures must exist around such progenitor 
stars. This prediction was rapidly borne out in observations of the
AGB star R Hya. Recent IR observations of the star, taken as part of the 
MIRIAD programme, revealed an arc-like structure to the North-West of 
the star \citep{ueta06}, aligned in the direction of the star's proper
motion. \cite{rhya} modelled this structure as an AGB wind-ISM bow shock 
ahead of the star, confirming the existence of such structures and 
showing that the dominant shaping factor for PN--ISM interaction and the 
origin of PNe rebrightening is formed during the preceding AGB phase of 
evolution. 

Evidence for similar bow shocks has now been found around several other 
giant and AGB stars, e.g. RX Lep \citep{libert08} and R Cas \citep{ueta09}. 
In the case of the famous red giant $\alpha$ Orionis, observations 
presented by \cite{noriega97} revealed an arc-like structure through
a novel data-reduction technique. The observations constitute what is 
probably the earliest evidence for a late-evolution stellar wind-ISM 
bow shock, recently confirmed and analytically modelled by 
\cite{ueta08b}. None of these cases, including that of R Hya, found
any clear evidence for a tail of ram-pressure-stripped material behind 
the bow shock.

\begin{figure}[ht]
\begin{center}
\includegraphics[width=7cm]{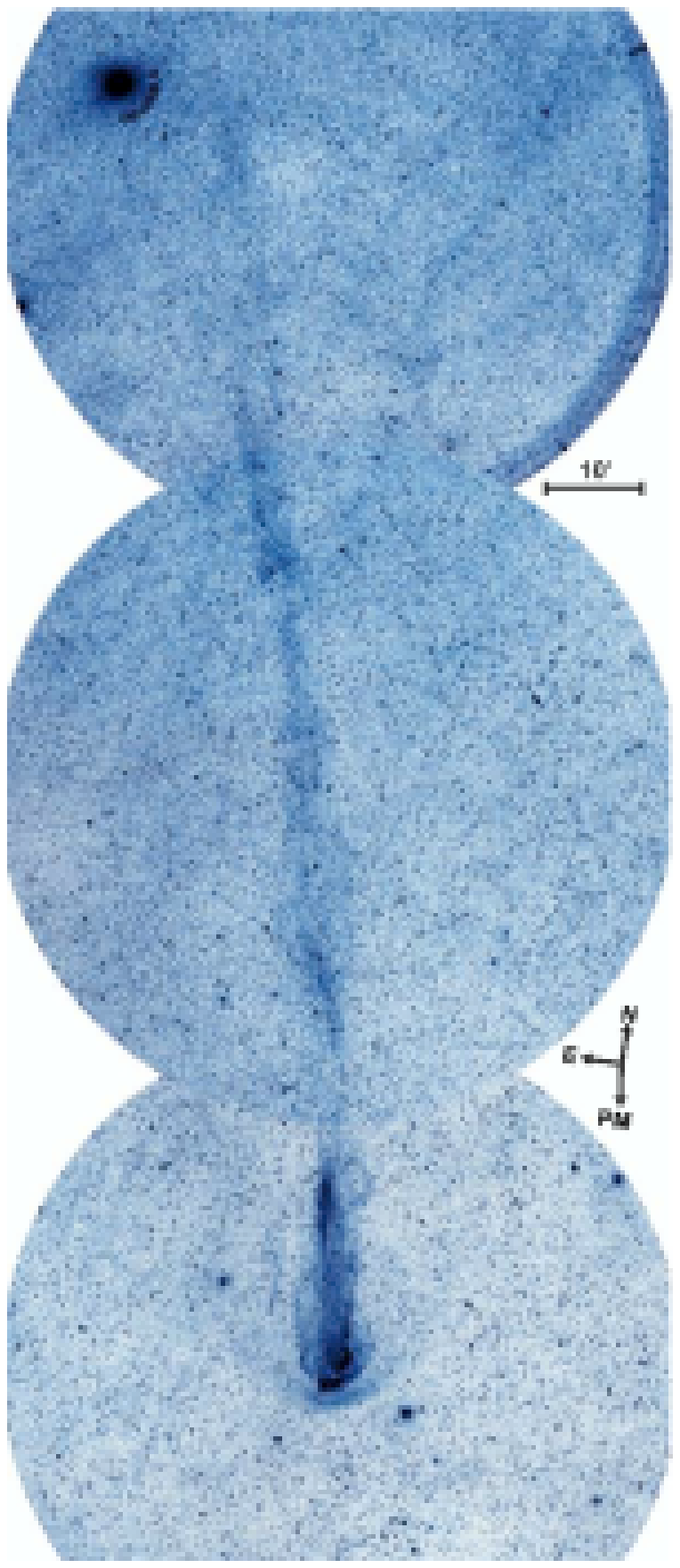}
\caption{A reproduction of figure 1b from \cite{mira} showing the 
2 degree structure stretching behind the location of the Mira binary 
system.}
\label{mira}
\end{center}
\end{figure}

Definitive evidence of such a tail was fortuitously discovered behind
the famous Mira binary system as part of routine GALEX UV observations, 
originating from the AGB star Mira A \citep{martin07}. The narrow collimated 
tail stretches 2 degrees on the sky, equivalent to 4 pc at the distance 
of the binary system, behind an arc-like structure ahead of the system. 
We reproduce an image of the structure in Figure \ref{mira}.
\cite{mira} modelled the structure as a bow shock and 
ram-pressure-stripped tail stretching away from the location of the 
binary system. For a full discussion of Mira's tail and its
characteristics, we review the reader to the recent review of
\cite{philtrans}. Close to the binary system, there is evidence for
asymmetric outflow close to the system \citep{meaburn09}. This recent 
outflow ($\sim1000$ years old), whilst not aligned with the tail, will
strongly affect the PN which Mira A will eventually form. The most-likely bipolar PN will
then later experience a complex rebrightening as it interacts with the bow
shock. Given the relatively high speed and close proximity of the 
bow shock to the binary system, the simulations of \cite{pn-ism} 
suggest Mira's PN will relatively rapidly interact with the ISM.
Mira and its tail are a template progenitor of PN--ISM interaction. 

\section{The progeny of rebrightened PNe: implications for future observational work}

The four stage classification of \cite{pn-ism} predicts the future of
rebrightened PNe in the final stage. At this time, their model 
predicts a deceleration of the
shell in the direction of motion combined with ram-pressure-stripping
of material, driving the majority of nebular material into the tail.
The fast wind from the CSPN has formed another bow shock, much closer
to the CSPN. The remaining AGB wind material and the remnants of the 
PN shell are swept downstream in turbulent regions of high density and 
temperature as illustrated in Figure \ref{pnstages}(d). Wareing et al.
note that any observable structure would be difficult to identify as a 
PN.

\begin{figure}[ht]
\begin{center}
\includegraphics[angle=0,angle=90,width=3.5cm]{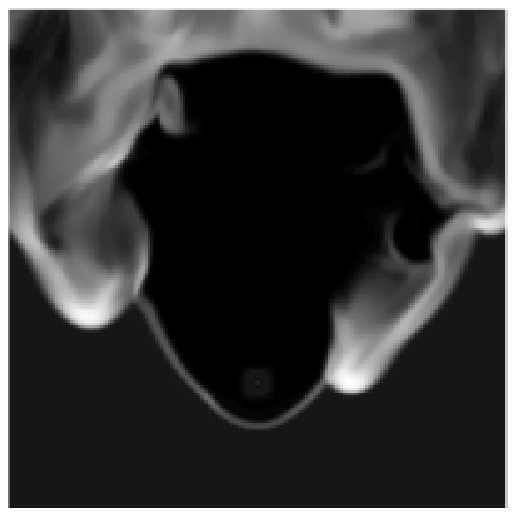}
\includegraphics[angle=0,angle=90,width=3.5cm]{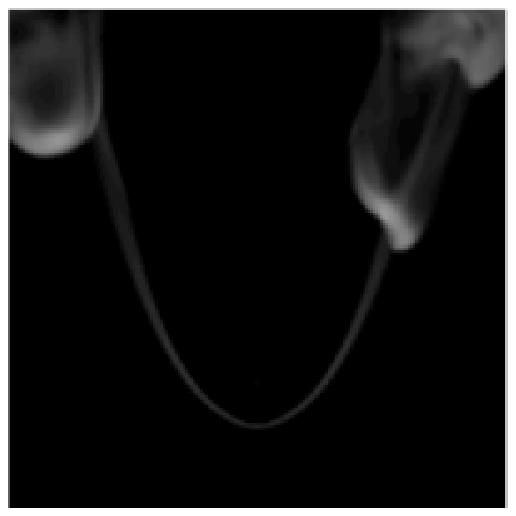}
\caption{Panels showing gas density during the PN phase for a PN moving
at a relative speed of 125 \kms. On the left we show a slice through 
the gas density datacube at the position of the central star, 
perpendicular to the direction of motion, at 15,000 years into the PN 
phase, on the right at 20,000 years into the PN phase. The slices are
1 pc on a side.}
\label{extremepn}
\end{center}
\end{figure}

We are able to show two snapshots of this evolution in Figure 
\ref{extremepn}. The snapshots are taken from the Case E simulation
presented in \cite{pn-ism}. They are at later times, 15,000 and 20,000
years into the PN phase, than the snapshots presented in Wareing et al.'s
work. Moving at 125 \kms\ through an average galactic plane ISM density 
of $n_{\rm H} = 2$ cm$^{-3}$, the PN has long since entered the fourth stage 
of interaction. The fast wind bow shock has formed against the oncoming 
ISM and the remains of the nebular shell and AGB wind bow shock are 
being swept downstream in turbulent regions of high density and 
temperature. 

\begin{figure}[ht]
\begin{center}
\includegraphics[angle=0,width=8.5cm]{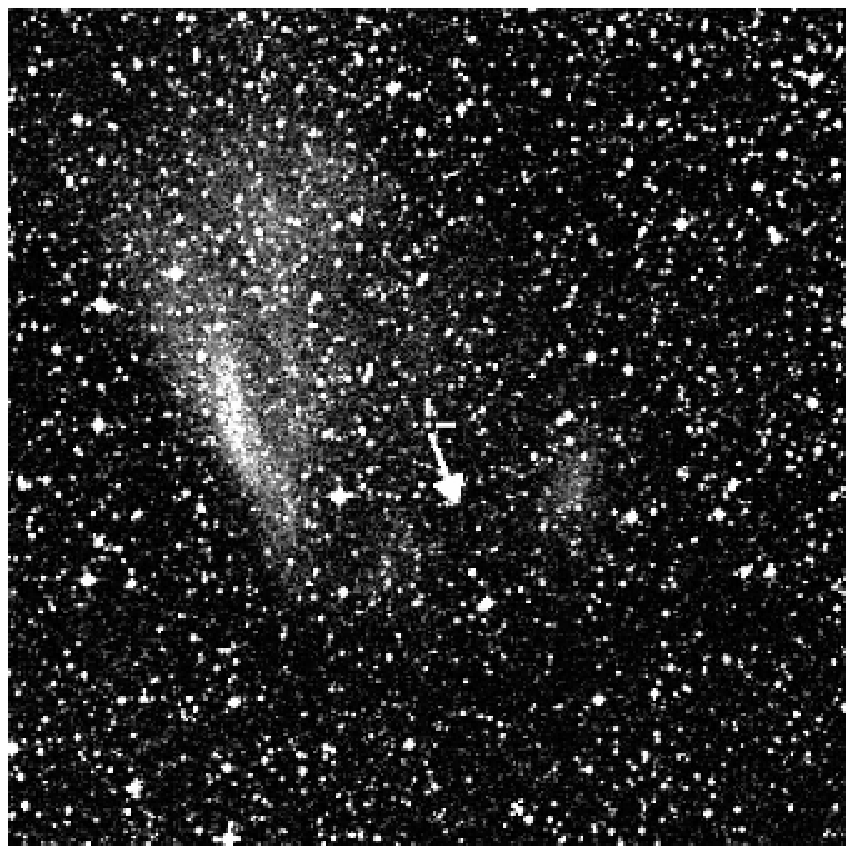}
\caption{An image of the PN Sh 2-68. The CSPN is indicated between
the markers and the arrow indicates the direction of the proper motion;
$202\deg \pm 6\deg$ East of North \cite{kerber02}. In the image, North is up
and East to the left. The image has been retrieved via Aladin from the 
DSS2, the ``Second Epoch Survey'' of the southern sky produced by the 
Anglo-Australian Observatory (AAO) using the UK Schmidt Telescope.}
\label{sh2-68}
\end{center}
\end{figure}

There are very few possible detections of such objects in the literature
and given their likely advanced age and high speed central stars, this
comes as no surprise.
The primary example is Sh 2-68, a nebula that shows all the signs of 
being swept downstream behind the central star. The CSPN which has one of the 
largest measured proper motions - $0.053 \pm 0.005 
{\rm arcsec}\ {\rm yr}^{-1}$ \cite{kerber02}. Various distance estimates
exist to the nebula, but range over 500-1000 pc \cite{napiwotzki91,napiwotzki95}.
At this distance, the CSPN is likely at the very high speed end of the
CSPN velocity range \citep{burton88}, around 150 \kms. The filamentary structure
of the nebula is attributed by Kerber et al. to a Rayleigh-Taylor
instability, which fits well with previous theoretical studies
\cite{dgani94,dgani98}. However, the filaments are all aligned with the 
direction of motion and they could simply be the rebrightened regions of high density
and temperature predicted by the four-stage model being ram-pressure-stripped into the
tail. In particular, the strongest emission from Sh 2-68 comes from
a long arc, at the edge of the emission and 'pointing' in the direction
of the CSPN. We show an image of the nebula in Figure \ref{sh2-68}.
This fits well with Wareing et al.'s structures shown in
Figure \ref{extremepn}. Note also that faint material ahead of the star
appears to be confined by the inferred position of the ancient bow shock.
Interestingly, \cite{xilouris96} found a extended halo around Sh 2-68
in the form of a cometary tail in opposition to the direction of 
proper motion. If this material
is connected to the nebula, and the alignment with the direction of
proper motion strongly supports this, then Sh 2-68 is the key example
of the fate of rebrightened PN and deserving of future investigation.

Large, faint objects such as Sh 2-68 and Sh 2-188 are now regularly being
observed as part of sky surveys, e.g. IPHAS, VPHAS, MASH-II, and new
astronomical tools are able to take deeper observations, e.g. the Herschel space telescope
will soon begin its survey of cold dust around evolved stars
and is likely to reveal many more tails such as the one behind Mira.
These surveys and instruments are likely to reveal the ancient, faint,
extended progenies of rebrightened PNe, such as Sh 2-68. The structures
will have little resemblance to young PN, their central stars may be far
from the strongest emission and any remnants of nebular shells are likely
to have been long since swept into turbulent regions in the tail. 
Identification of candidate CSPN will be possible by proper motion studies
- elements of the structure are highly likely to be a aligned with the
direction of motion. These objects will have to be carefully filtered
and confirmations of a candidate stage WZO 4 PN revealed by IPHAS are 
already proving difficult (private communication: L. Sabin; see
Sabin's paper elsewhere in this volume).

\section{Concluding remarks}

Simulations have shown that interaction with the ISM strongly affects 
the death of a PN, even one with an average velocity central star. The
structure of the AGB wind around the CSPN is changed from a wall into a
bow shock during the preceding phase of evolution. This bow shock and
the ram pressure stripping of material into the accompanying tail then
dominates the evolution of the PN. Rebrightening and rejuvenation occur,
on a short timescale for high speed CSPN, as the PN shell interacts with 
the bow shock. Following this, the PN is stripped downstream in turbulent
regions moving down the tail, resulting in the central star leaving its 
PN.

Further study of rebrightened PN and their surroundings can provide
considerable information regarding several other research avenues.
As highlighted in the opening paragraph, tracing the ejected mass in
circumstellar structures such as bow shocks, tails and late-evolution
turbulent regions swept down the tail can reveal much about the way
AGB and postAGB stars lose mass. Such information as the position of
the bow shock and the parameters of the tail can also reveal, taking
into account the method of formation, not only historical stellar mass-loss rates and local ISM conditions,
but also details of galactic orbits for the first time. The few objects 
that have observed lengthy tails, including 
Mira, HFG1 and Sh 2-68, are currently uniquely placed to derive such
orbits. Future surveys of cool dust around AGB stars should reveal many more 
tails allowing a deeper investigation of a meaningful sample.

Once we have a meaningful sample of rebrightened PNe, we should also
be able to see how the observable lifetimes of PNe are extended. Current
statistical population estimates, which depend on estimated lifetimes, 
indicate a larger population than we have currently detected, but by 
extending that lifetime, population estimates accordingly reduce and 
we are likely to see much more of an agreement between the numbers we 
actually observe, which is believed to be fairly complete within the Milky Way, and the predicted population.

The model of PN--ISM interaction then can account for many characteristics
of PNe and explain highly complex, disrupted objects such as Sh 2-68, but
it does not reproduce several observed details. In particular,
the fragmentation of the nebular rim is not reproduced by the simulations.
More advanced models must be able to reproduce this, at least in the cases
of high speed CSPN as it seems common at this end of the velocity 
distribution, e.g. Sh 2-188. No theoretical investigation has yet been
performed employing 3D MHD simulations which include the galactic magnetic
field. This would seem a natural next step and as Dgani and Soker point
out, this may rapidly provide an explanation of particularly fragmentary
shells. In a greater challenge to the model, the existence of ancient, effectively
spherical PNe with very few signs of ISM interaction are difficult to explain
within the model context, especially if they have displaced
central stars which rules out an alignment between viewing angle and direction
of motion. The model of PN--ISM interaction has come along way and revealed
the phenomenon of rebrightened PNe, but there are a still a number of open questions
to be solved and no doubt future observations will reveal many more.

\section*{Acknowledgments} 

The author acknowledges and thanks several researchers who's input has
greatly helped in this project and the creation of this review paper
including Prof. A. A. Zijlstra, Ass. Prof. O. De Marco, Dr T. J. 
O'Brien, Dr M. Lloyd and Dr L. Sabin. CJW is funded through a 
post-doctoral Research Fellowship supported by the Science \& Technology 
Facilities Council [grant number PP/E001092/1].


\end{document}